# Cascade of Even-Denominator Fractional Quantum Hall States in Mixed-Stacked Multilayer Graphene

Yating Sha[1,2,†], Kai Liu[1,2,†], Chenxin Jiang[3,4,5,†], Dan Ye[3], Shuhan Liu[2], Zhongxun Guo[6], Jingjing Gao[6], Ming Tian[7], Neng Wan[6,7], Kenji Watanabe[8], Takashi Taniguchi[9], Bingbing Tong[10,11], Guangtong Liu[10,11]\*, Li Lu[10,11], Yuanbo Zhang[6], Zhiwen Shi[2], Zixiang Hu[3,4]\*, Guorui Chen[1,2]\*

[1]*Key Laboratory of Artificial Structures and Quantum Control (Ministry of Education), School of Physics and Astronomy, Shanghai Jiao Tong University, Shanghai 200240, China*

[2]*Tsung-Dao Lee Institute, Shanghai Jiao Tong University, Shanghai 201210, China*

[3]*Department of Physics, Chongqing University, Chongqing 401331, China*

[4]*Chongqing Key Laboratory for Strongly Coupled Physics, Chongqing University, Chongqing 401331, China*

[5]*Division of Physics and Applied Physics, Nanyang Technological University, Singapore 637371*

[6]*State Key Laboratory of Surface Physics and Department of Physics, Fudan University, Shanghai 200433, China*

[7]*Key Laboratory of MEMS of Ministry of Education, School of Integrated Circuits, Southeast University, Jiangsu, Nanjing 210096, China*

[8]*Research Center for Electronic and Optical Materials, National Institute for Materials Science, 1-1 Namiki, Tsukuba, Japan*

[9]*Research Center for Materials Nanoarchitectonics, National Institute for Materials Science, 1-1 Namiki, Tsukuba, Japan*

[10]*Beijing National Laboratory for Condensed Matter Physics, and Institute of Physics, Chinese Academy of Sciences, Beijing 100190, China*

[11]*Hefei National Laboratory, Hefei, Anhui 230088, China*

[†]These authors contributed equally to this work.

\*Correspondence to: chenguorui@sjtu.edu.cn, zxhu@cqu.edu.cn, gtliu@iphy.ac.cn

The fractional quantum Hall effect (FQHE), particularly at half-filling of Landau levels, provides a unique window into topological phases hosting non-Abelian excitations. However, experimental platforms simultaneously offering large energy gaps, delicate tunability, and robust non-Abelian signatures remain scarce. Here, we report the observation of a cascade of even-denominator FQH states at filling factors $v = -5/2, -7/2, -9/2, -11/2$ and $-13/2$, alongside numerous odd-denominator states in mixed-stacked pentalayer graphene—a previously unexplored system characterized by intertwined quadratic and cubic band dispersions. These even-denominator states, representing the highest filling half-filled states reported so far in the zeroth Landau level (ZLL), emerge from two distinct intra-ZLL and exhibit unprecedented displacement-field tunability driven by LL crossings in the hybridized multiband structure. At half fillings, continuous quasiparticle phase transitions between paired FQH states, magnetic Bloch states, and composite Fermi liquids are clearly identified upon tuning external fields. Numerical calculations, revealing characteristic sixfold ground-state degeneracy and chiral graviton spectral analysis, suggest the observed even-denominator FQH states belong to the non-Abelian Moore-Read type. These results establish mixed-stacked multilayer graphene as a rich and versatile crystalline platform for exploring tunable correlated topological phases.

**Main text**

Fractional quantum Hall (FQH) states are collective two-dimensional electron phases arising under strong magnetic fields when Landau levels (LLs) are partially filled[1]. Most known FQH states occur at fillings with odd denominators (e.g., 1/3, 2/5) and can be captured by the composite fermion (CF) theory as integer quantum Hall states of bound electron–flux quasiparticles[2]. A long-standing exception is the even-denominator FQH state at filling $v = 5/2$ in the second LL (orbital quantum number $N = 1$) of GaAs[3,4]. Several candidate states have been proposed to explain the 5/2 state[5–7], most notably the Moore–Read Pfaffian[8], its particle-hole conjugate (anti-Pfaffian)[9,10], and a particle–hole symmetric (PH-Pfaffian) state[11], all featuring $p$-wave paired CFs[12] and hosting non-Abelian quasiparticles that are central for topological quantum computation[7,13]. Despite decades of study, unambiguous experimental proof of the non-Abelian nature of the 5/2 state remains elusive[14,15]. One major challenge is the fragility and limited tunability of even-denominator states in conventional GaAs heterostructures[16].

Recent advances in van der Waals materials, particularly graphene-based systems, have opened new avenues for exploring FQH states. In particular, multilayer graphene can host complex Landau level spectra with tunable degeneracies[17,18]. Bilayer graphene was predicted to support a Pfaffian-like half-filled state in its $N = 1$ LL[19], and experimental evidence of the 5/2 state has been observed in ultra-clean bilayer graphene devices[20–23]. More recently, even-denominator states have been reported in trilayer graphene[24] and other multicomponent quantum Hall systems[25,26]. Multilayer graphene offers unique advantages: the large degeneracy of the ZLL in multilayer graphene promotes a variety of spontaneously broken symmetries, producing conditions for exotic particles. Recently, non-Abelian FQH states are predicted at both odd and even filling fractions of the highly degenerated LLs in rhombohedral multilayer graphene[27,28]. Moreover, the orbital composition of LL wavefunctions can be controlled by an external perpendicular electric field (displacement field $D$)[29,30]. By tuning $D$, one can mix LL orbitals and adjust Coulomb interactions[21,30,31], potentially stabilizing Moore–Read states. This tunability has already enabled observations of sequences of even- and odd-denominator states and related phase transitions in Bernal-stacked multilayer graphene[21,24].

ABCBC-stacked pentalayer graphene (ABCBC-5LG), one of the mixed-stacking sequence with non-centrosymmetric lattice[32–35] (Extended Data Fig. 3), can be regarded as a stack of an ABC-trilayer and an AB-bilayer graphene sheets (Fig.1b). The ABCBC stacking sequence inherently lacks inversion and mirror symmetry due to the inequivalent chemical environments of the carbon atoms, creating built-in displacement fields. As a result, its band structure combines cubic and parabolic band dispersions near the Fermi level with a small intrinsic gap (Fig.1c, left), and undergoes relative energy shift between the cubic and parabolic bands under external perpendicular displacement field $D$ (Fig.1c, right). The basic transport behavior of ABCBC multiband has been identified in our recent work[36]. This hybrid double layer structure creates a hierarchy of ABC- and AB-graphene's degenerated Landau levels (LLs) with distinct orbital components, enabling LL crossings under electric or magnetic fields[18,37]. Unlike simpler graphene systems, the multi-band nature of ABCBC-5LG amplifies LL mixing effects, which are critical for stabilizing correlated states at fractional fillings. Moreover, dual electrostatic gating (see Methods) allows independent control over carrier density ($n$) and interlayer polarization ($D$), offering powerful experimental knobs to control phase transitions between competing orders.

In this work, we leverage these properties to uncover a rich spectrum of FQH states in ABCBC-5LG. By applying high $B$-field and $D$-field, we reveal an unprecedented cascade of even-denominator FQH states at half-filling of certain two intra-ZLLs, interwoven with conventional odd-denominator sequences. We demonstrate that the electric displacement field $D$ can drive transitions between odd-denominator Jain sequences and even-denominator Moore–Read states by inducing Landau level crossings. Among the even-denominator states, two record-high filling factors, -11/2 and −13/2, for ZLL are reported. Those states coexist with odd-denominator sequences that bifurcate into hierarchies of 2- and 4-flux CFs, reflecting the interplay between multi-orbital LLs and interactions. Our device also showcases continuous quantum phase transitions including candidate non-Abelian and Abelian FQH states, as well as magnetic Bloch state. These results establish mixed-stacked graphene, for the first time, as a promising platform for exploring exotic correlated and topological state, which can host flat band features similar to rhombohedral graphene[38–42], while the hybridization between subbands may lead to even richer states.

**Phase diagram in high magnetic field**

Our devices consist of ABCBC-5LG encapsulated in hexagonal boron nitride (hBN) and graphite top and bottom gate electrodes, and contacted via one-dimensional edge metal contacts[43]. Data in the main text are all from sample S1 (Fig. 1a, Extended Data Fig. 1 and 2). Under application of a high magnetic field ($B$ = 18 T), the degeneracy of Landau levels is fully lifted, yielding robust integer quantum Hall (IQH) gaps from $v$ = -1 to -8 and a rich spectrum of LLs that can cross and hybridize as a function of $D$. Figure 1d shows the longitudinal resistance $R_{xx}$ measured as a function of filling factor $v$ ($v = nh/(eB)$, $n$ is carrier density; $h$ is Planck's constant; $e$ is the electron charge) and displacement field $D$ at base temperature ($T$ = 14 mK) and $B$ = 18 T. Black regions indicate $R_{xx}$ minima, corresponding to robust quantum Hall states. In particular, we observe pronounced minima at numerous fractional fillings, including both odd- and even-denominator fractions up to eight intra-ZLLs (limited by gate leakage). We label the observed integer and fractional quantum Hall states in the lower panel of Fig. 1d, where the IQH states are labeled by grey regions. Notably, even-denominator FQH states, including $v$ = -5/2, -7/2, -9/2, -11/2, and -13/2, are highlighted by red lines. They are flanked by nearby odd-denominator states (blue and orange lines) belonging to Jain sequences with two or four flux quanta (CF$_2$ and CF$_4$).

The emergence of these half-filled states is strongly tied to Landau level crossings. As $D$ is varied, the orbital character of LLs shifts, and we observe enhanced $R_{xx}$ at certain integer $v$ where LL crossings occur. These crossings trigger phase transitions between different FQH sequences: on one side of a crossing, odd-denominator Jain-series states dominate, whereas on the other side, the even-denominator half-filled state emerges as a robust gap. In Fig. 1d, such transitions are evident as the system switches from, e.g., a sequence of $CF_2$ states (blue lines in the sketch) to the appearance of an incompressible state at half-filling (red lines) and suppression of other fractions. This indicates that tuning $D$ can stabilize the Moore–Read type candidates by pushing the system into a regime of optimal LL mixing and interaction.

Away from the special half-filling conditions, the Jain CF hierarchy is largely intact, demonstrating the high quality of our device. Figure 2a and 2b show linecuts of $R_{xx}$ (and corresponding Hall conductivity $\sigma_{xy}$) at fixed $D = +0.4$ V/nm for filling ranges $-1 \geq \nu \geq 0$ and $-2 \geq \nu \geq -1$. In these regimes, where a single LL branch (orbital $N = 0$) is active, we resolve a standard sequence of $CF_2$ states with -1/3, -2/3, -2/5, -3/5, -3/7, -4/7, -4/9, -5/9, -5/11, -6/11, and -4/3, -5/3, -7/5, 8/5, -10/7, -11/7, -13/9, -14/9, -16/11, -17/11 (blue region in Fig. 2, following the standard fillings $\nu = p/(2p\pm1)$ with $p$ is integer). We also observe several states consistent with $CF_4$ (four-flux) at higher fillings, including -5/7, -9/5, -9/7, -12/7, -14/11, and -19/11 (yellow region, following $\nu = p/(4p\pm1)$), though these are weaker. All these odd-denominator states show quantized Hall plateaus (in $\sigma_{xy}$) and thermally activated behavior in $R_{xx}$. Figures 2c and 2d (upper panels) present the temperature dependence of $R_{xx}$ for representative states; the activation gaps extracted (lower panels) from Arrhenius fitting follow the expected linear trend, consistent with CF theory and suggesting a $CF_2$ liquid at the half-filling. From a linear fit $\Delta = \hbar eB_{eff}/m_{CF}$ (where $B_{eff}$ is the effective magnetic field seen by CFs), we estimate an effective CF mass $m_{CF} = 0.785\ m_e$ (where $m_e$ is the free electron mass), comparable to values previously reported in GaAs/AlGaAs[44]. The extrapolated gaps approaching half-filling (e.g., $\nu = -1/2$ and $-3/2$) tend to be negative, attributed to the disorder-induced broadening of the CF LLs[44–46]. The intercept of the linear fit yields a broadening parameters $\Gamma \sim 4$ K, even smaller than that previously reported in Corbino[47] and graphite gate defined graphene[48].

**Two groups of even-denominator FQHs**

Focusing on the half-filled states, Fig. 3a–f provide detailed linecuts of $R_{xx}$ and $\sigma_{xy}$ in the vicinity of each even-denominator fraction corresponding to dashed lines in

Fig.1d. We identify two groups of half-filled states, each associated with a different intra-ZLL: one group includes $v$ = -5/2, -7/2 and -2/9 (Fig. 3a–c, purple shading), and the other includes $v$ = -9/2, -11/2, and -13/2 (Fig. 3d–f, pink shading). In each case, the even-denominator state appears as a pronounced $R_{xx}$ minimum, while nearby Jain odd-denominator states are suppressed[3]. This indicates that the system favors the possible paired CF state over the sequence of single CF states in that filling range. By tracing the evolution of the LLs under $D$-tuning, we confirmed that the purple group (−5/2, −7/2, -9/2) and red group (−9/2, −11/2, −13/2) correspond to two distinct intra-ZLLs crossing the Fermi level. Starting from the states at -5/2 ($D$ = 0.4 V/nm) and -9/2 ($D$ = 0.24 V/nm), each time a LL crossing occurs as $D$ increases, the corresponding even-denominator state moves to a higher-index LL, following an upward-left trend (red/purple shading in Fig. 1d). At fixed half-filling of a certain LL, such as the fifth LL, by tuning $D$, the ground state can be continuously switched from FQH gapped state (-9/2) to composite Fermi liquid (CFL)[49] to FQH gapped state again. Each switch corresponds to a LL crossing. This behavior underscores a key result: the half-filled FQH state can repeatedly emerge in the same or different LLs of the same device, given appropriate tuning, forming a cascade of even-denominator FQH states. It is remarkable that we observe the sequence up to $v$ = -11/2 and -13/2, the highest half-filling even-denominator FQH states in ZLL reported so far, indicating the robustness afforded by the multilayer graphene platform.

To elucidate the nature of the observed even-denominator FQH states, we performed numerical calculations that incorporate three key physical effects: Coulomb interactions, LL mixing and $D$. The results reveal two distinct valence-band LLs with particularly large contributions from the $N$ = 1 LL (see Method and SM), naturally accounting for the two groups of half-filling FQH states observed in our experiment. In our analysis, we consider the mixing of two adjacent relativistic LLs, which leads to two types of fillings: 1+1/2 and 1/2 (see SM). Accordingly, the experimentally observed filling factors can be expressed as: $v$ = -4+3/2, -4+1/2, -6+3/2, -6+1/2 and -8+3/2 as shown in Fig. 3b. All of these cases exhibit six-fold quasi-degenerate ground states in our numerical calculations—a hallmark of the Moore-Read type state. To gain deeper insight into their particle-hole symmetry, we compute the chiral graviton spectral functions. By comparing the spectral responses, we find that the states at -5/2, -9/2, -13/2 tend toward the anti-Pfaffian phase. In contrast, the states at -7/2, -11/2 exhibit characteristics more consistent with the Pfaffian phase. Signatures of the daughter states[49] associated with -5/2 and -9/2 are shown in Extended Data Fig. 8. Together, these results reveal a previously unexplored cascade of even-denominator FQH states, shaped by band structure and LL hybridization.

**Phase transition between distinct quasiparticles**

To map out the LL structure more comprehensively, we also performed measurements as a function of magnetic field $B$ at fixed displacement field. Figure 4a presents $R_{xx}$ vs $v$ and $B$ at $D = 0.4$ V/nm, focusing on the region where LL crossings occur. As $B$ is varied, we observe some dark-colored diagonal traces (representing $R_{xx}$ dips) occur accompanied by some bright-colored squares-shaped regions (representing large $R_{xx}$). These features relate to a Hofstadter butterfly pattern—oscillations in $R_{xx}$ reflecting the fractal LL spectrum formed in a periodic moiré potential between graphene and hBN[50]. We find that these patterns are observed solely near LL crossings. This can be attributed to the relatively weak moiré potential in this device (moiré period ≈ 12.8 nm; see Extended Data Fig. 4), leading to small Hofstadter gaps that are only visible when comparable to the Landau level gap, that is, near LL crossings. Moreover, our second ABCBC-5LG device without moiré also reveals a consistent phase diagram in high-field, confirming the observed states is intrinsic to the special stacking configuration (Extended Data Fig. 5).

Within this region, the appearance and disappearance of even-denominator states can be tracked: at certain $B$ values near 16-18 T, the $v = $ -5/2 or -7/2 FQH states exists. While decreasing $B$ field, the FQH states disappear and the LLs are transitioned into Hofstadter states first, and then odd-denominator FQH and CFL states emerge. Notably, from Fig. 4a, the tunable transitions at half-filling are obvious—from even-denominator fractional quantum Hall (FQH) states, through Hofstadter miniband regimes, to CFL phases—suggest a sequence of continuous quasiparticle phase transitions. As the magnetic field is varied, the nature of the low-energy excitations evolves: from a gapped state that may involve paired CFs, to magnetic Bloch electrons, and eventually to a gapless CFL with an emergent Fermi surface (Fig. 4b). This progression reflects distinct changes in the emergent quasiparticles, shaped by the interplay between interactions, magnetic flux, and lattice effects in the multilayer graphene system.

Interestingly, we observe a qualitative difference in the nature of these transitions at half-filling between different phases. Figure 4b shows linecuts of Fig. 4a at $v = -7/2$ and -5/2 along the magnetic field, where the transition from the CFL (blue shaded) to the magnetic Bloch state (yellow shaded) appears relatively gradual, with a slow variation in longitudinal resistance across the transition region. In contrast, the transition from the gapped even-denominator FQH state (purple shaded)—presumably a paired CF phase—to the Bloch regime is significantly sharper, as indicated by a steep change in $R_{xx}$. This contrast suggests that the two transitions may be captured by distinct quantum phase transitions scenarios. The gradual CFL-to-Bloch transition likely involves a reconstruction of the Fermi surface in response to a periodic potential, possibly without a change in topological order. In contrast, the sharp breakdown of the paired FQH state may indicate a topological phase transition, wherein a non-Abelian incompressible fluid gives way to a topologically trivial band insulator. The differing sharpness may thus reflect the presence or absence of a bulk energy gap and the collapse of long-range entanglement across the transition. The underlying mechanism deserves further investigations.

**Methods**

**Sample fabrications:** Graphene, graphite and hBN are mechanically exfoliated on SiO$_2$(285nm)/Si substrates, and the layer numbers are identified using optical contrast and atomic force microscopy. The stacking order of pentalayer graphene is identified using scanning near-field optical microscopy (SNOM), where different stacking orders have different contrast.

The heterostructures were assembled using a dry-transfer technique with polypropylene carbonate (PPC) or polycarbonate (PC) as the polymer, resulting in hBN/top graphite/5LG/hBN/bottom graphite stacks. The stacks were annealed to remove interfacial contamination and trapped bubbles. SNOM imaging was then performed again to verify the special stacking order.

Subsequently, devices were patterned into Hall bar geometries using standard electron-beam lithography, reactive ion etching (using CHF$_3$/O$_2$ plasma, O$_2$ plasma or SF$_6$ plasma as needed, depending on the target material), and one-dimensional Cr/Au (about 5/50 nm) edge-contact deposition.

**Transport measurements:** Transport measurements were carried out in a dilution refrigerator (base temperature ~ 14 mK), equipped with superconducting magnets up to 18 T. Standard low-frequency lock-in techniques were used with an AC excitation current of 1–10 nA at ~17.777 Hz. Home-built low-pass RC filters were installed in the fridge to suppress high-frequency noise.

The data of Hall conductivity $\sigma_{xy}$ are obtained from the measured resistances by $\sigma_{xy} = \rho_{xy}/(\rho^2_{xx} + \rho^2_{xy})$. Gate voltages were applied using two Keithley 2400 source meters. Independent top and bottom graphite gates enabled tuning of the carrier density $n$ and the out-of-plane displacement field $D$ in the graphene layer. Based on a parallel-plate capacitor model: $D = (D_b + D_t)/2$, $n = (D_b - D_t)/e$, where $D_b = +\varepsilon_b(V_b - V_b^0)/d_b$, $D_t = -\varepsilon_t(V_t - V_t^0)/d_t$, $\varepsilon$ and $d$ are the dielectric constant and thickness of the dielectric layers, respectively, $V_b^0$ and $V_t^0$ are effective offset voltages caused by environment-induced carrier doping. Notably, in such devices, portions of the graphene leads unavoidably lie outside the dual-gated region. To independently control the carrier density in these ungated regions, we employed the silicon substrate as a global back gate to lower contact resistance and improves the quality of magnetotransport measurements.

**Calculation of CF effective mass:** A very fundamental parameter characterizing CFs is their effective mass ($m_{CF}$), which arises primarily from electron-electron interactions. The magnitude of this mass determines the energy separation between the CF LLs (sometimes referred to as Λ "levels"), and in turn determines the size of the energy gaps for the FQHSs at $v = p/(2p+1)$. To extract $m_{CF}$, we measured the temperature dependence of the longitudinal resistance $R_{xx}$ for a series of well-developed odd-denominator FQHSs. Activation gaps $\Delta$ were obtained by fitting the data to the Arrhenius form $R_{xx} \propto \exp(-\Delta/2K_BT)$, using linear fits of $\ln(R_{xx})$ versus $1/T$ in the thermally activated regime. Within the CF framework, these gaps correspond to the CF cyclotron energy $\hbar\omega_{CF} = \hbar eB_{eff}/m_{CF}$, where $B_{eff} = B - B_{half\_filling}$ is the effective magnetic field seen by CFs.

## Acknowledgments

G.C. acknowledges support from NSF of China (grant nos. 12350005 and 12174248), National Key Research Program of China (grant nos. 2021YFA1400100 and 2020YFA0309000), Shanghai Science and Technology Innovation Action Plan (grant no. 24LZ1401100), and Yangyang Development Fund. Y.S. and K.L. acknowledge support from T.D. Lee scholarship. K.L. acknowledges support from the National Natural Science Foundation of China (grant no. 124B2071). Z.H. acknowledges support from NSF of China (grant nos. 12474140 and 12347101). C.J. acknowledges the support of the China Scholarship Council (grant no. 202406050101) and thanks B.Y. and Y.W. for fruitful discussion. N.W. acknowledges support from Open Project of Key Laboratory of Artificial Structures and Quantum Control (Ministry of Education), Shanghai Jiao Tong University. K.W. and T.T. acknowledge support from the JSPS

KAKENHI (Grant Numbers 21H05233 and 23H02052), the CREST (JPMJCR24A5), JST and World Premier International Research Center Initiative (WPI), MEXT, Japan. Y.Z. acknowledges support from National Key R&D Program of China (grant no. 2022YFA1403301), and New Cornerstone Science Foundation. Z.S. acknowledges support from the National Key R&D Program of China (No. 2021YFA1202902 and No. 2022YFA1402702 and the National Science Foundation of China (No. 12374292). This work was carried out at the Synergetic Extreme Condition User Facility (SECUF, https://cstr.cn/31123.02.SECUF).

**Competing interests**

The authors declare no competing financial interests. Data availability All data that support the findings of this study are available from the contact author upon request.

**Data availability**

All data that support the findings of this study are available from the contact author upon request.

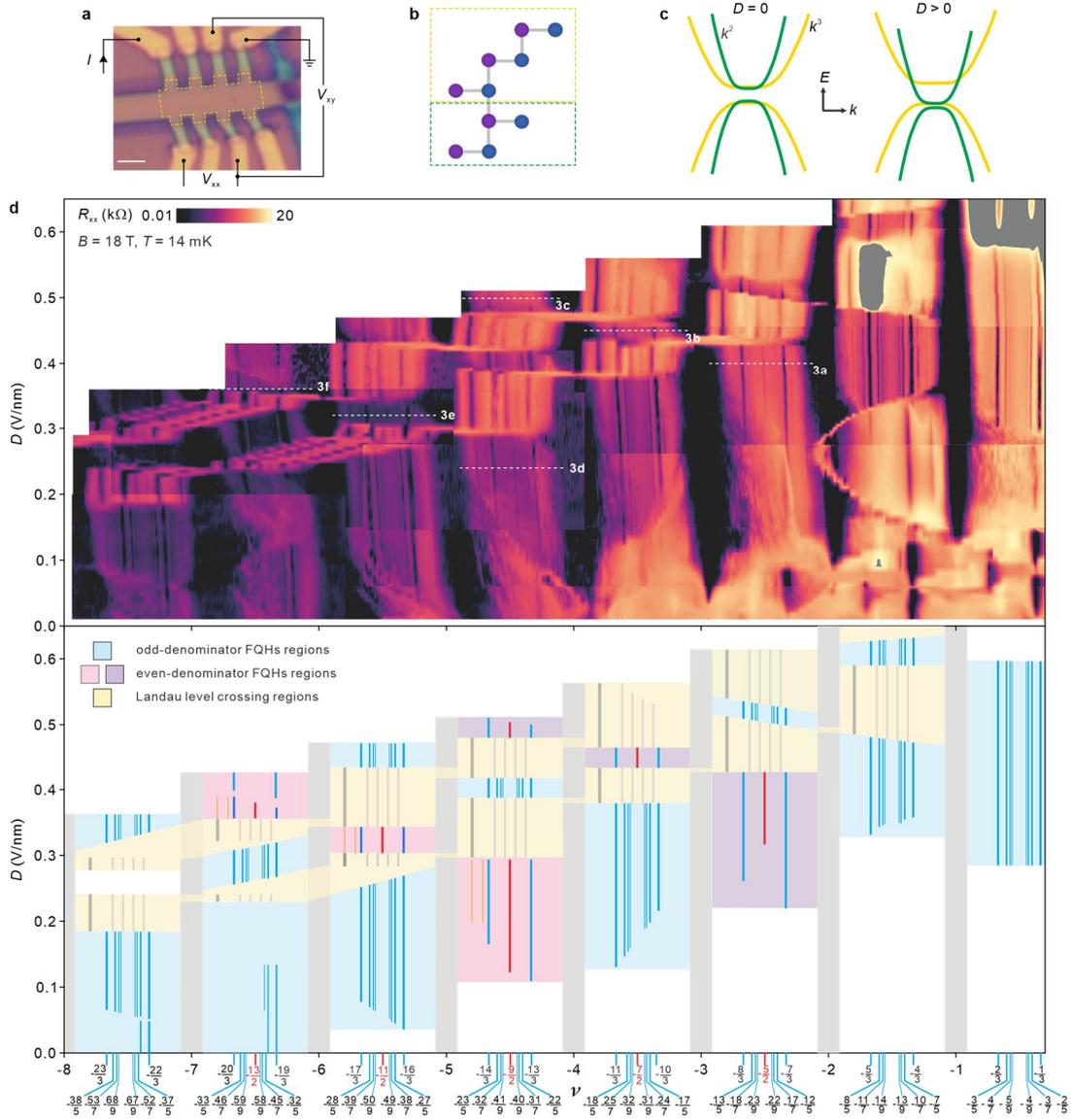

**Fig. 1 | Cascade of FQH states in ABCBC-5LG under large +*D*. a,** Optical image of the graphite/hBN/ABCBC-5LG/hBN/graphite device, - a schematic of the transport measurement configuration. The Hall bar shaped graphene is outlined by the yellow dashed line. Scale bar: 2 μm. **b,** Stacking configuration of ABCBC-5LG, which can be conceptually understood as a composition of an AB-2LG (outlined in the green box) and an ABC-3LG graphene (outlined in the yellow box). **c,** Schematic band structure of ABCBC-5LG at $D = 0$ and $D > 0$. The parabolic bands (green) and cubic bands (yellow) originate from the bilayer and trilayer parts, respectively, as shown in **b**. The hybridization between the two bands is neglected for simplicity. **d,** $R_{xx}$ as a function of filling factor $v$ and displacement field $D$ at $T = 14$ mK and $B = 18$ T. Well-resolved $R_{xx}$ minima are observed at fractional fillings with both even- and odd-denominators. Landau level crossings are characterized by increased $R_{xx}$ at integer fillings, followed by phase transitions between odd- and even-denominator sequences of FQH states. Bottom panel: sketch map of the top panel. Grey shaded regions correspond to the integer quantum Hall states. Blue and orange lines label the odd-denominator FQH states of two-flux ($CF_2$) and four-flux ($CF_4$) composite fermions. Red lines label the even-denominator FQH states.

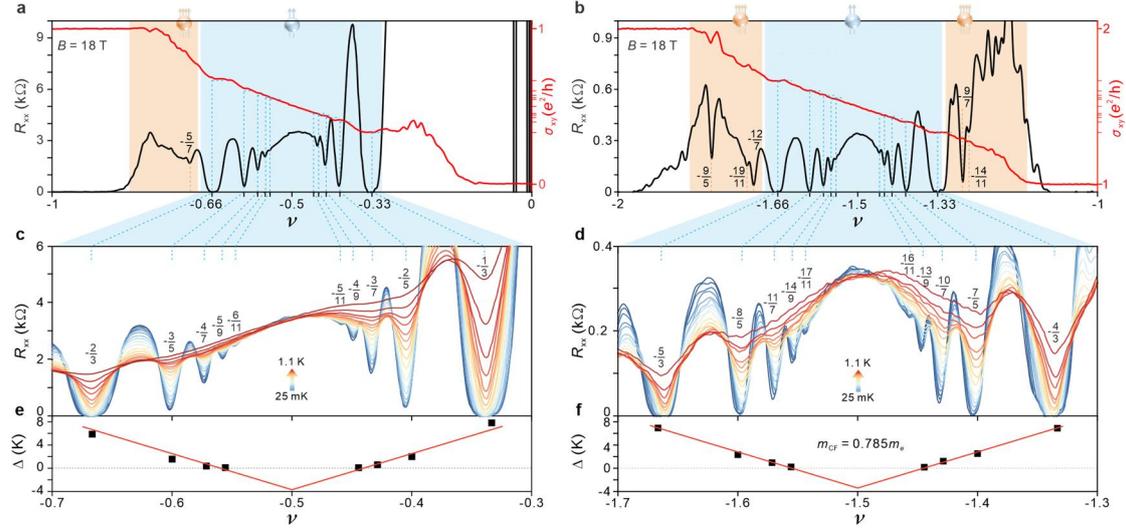

**Fig. 2 | Energy gaps of odd-denominator FQH states and the effective mass of the CF$_2$ quasiparticles. a-b,** Linecuts of $R_{xx}$ and $\sigma_{xy}$ for the FQH states within the filling ranges of $-1 \leqslant \nu \leqslant 0$ and $-2 \leqslant \nu \leqslant -1$, respectively, measured at $B = 18$ T and $D = +0.4$ V/nm. The Jain sequence of two-flux CFs (CF$_2$), is evident in the blue region, while several four-flux CFs (CF$_4$) states are also observed in the orange shading region. **c-d,** Temperature dependance of $R_{xx}$ for the sequence of CF$_2$ states shown in **a** and **b** respectively. **e-f,** Corresponding energy gaps for various FQH states, determined from Arrhenius fitting (Extended Data Fig. 7). The gaps follow a linear fit (red straight line in the lower panels): $\Delta = \hbar eB_{eff}/m_{CF}$, where $B_{eff}$ is the effective magnetic field for CFs and $m_{CF}$ is composite fermion mass. $m_{CF}$ is found to be $0.785m_e$.

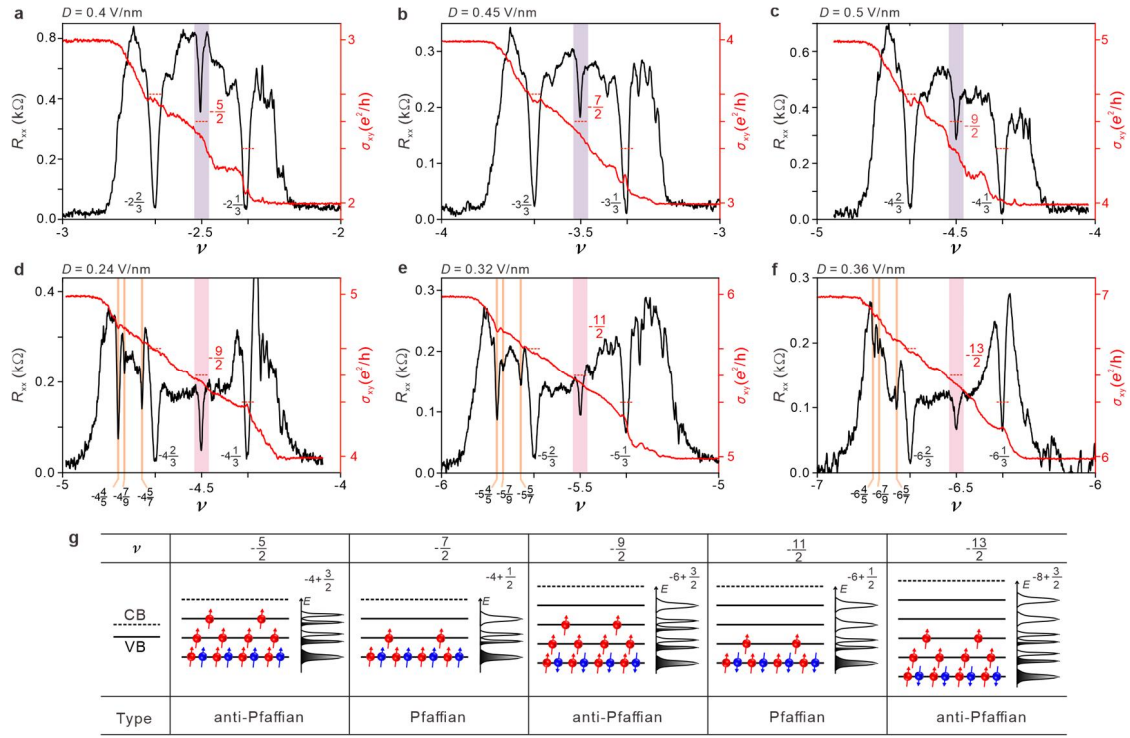

**Fig. 3 | Two sets of even-denominator FQH states. a-f,** Linecuts of $R_{xx}$ and $\sigma_{xy}$ at $B = 18$ T within the filling ranges of $-3 \leqslant \nu \leqslant -2$ (**a**), $-4 \leqslant \nu \leqslant -3$ (**b**), $-5 \leqslant \nu \leqslant -4$ (**c** and **d**), $-6 \leqslant \nu \leqslant -5$ (**e**) and $-7 \leqslant \nu \leqslant -6$ (**f**), respectively. The displacement field $D$ for each panel is indicated by white dashed lines in Fig.1d. Even-denominator FQH states at $\nu = -5/2, -7/2, -9/2, -11/2$ and $-13/2$ are clearly resolved, accompanied by a strong suppression of odd-denominator Jain states. We categorize the observed half-filling states into two groups, highlighted by purple shading in (**a-c**) and pink shading in (**d-e**). Phenomenologically, a hallmark of the second group is the appearance of several CF$_4$ states (orange shading) accompanying each half-filled state on the left, in contrast to the first group where such states are absent. Similar features reproducibly observed at opposite field direction, namely at $-18$ T (Extended Data Fig. 6). **(g)** Identification of the Moore-Read type for all observed half-filled states based on the calculations (see SM), along with schematic illustrations of the relevant fillings. Solid and dashed lines represent LLs in the valance and conduction band, respectively. Red and blue sphere denote electrons, with arrows indicating their spin.

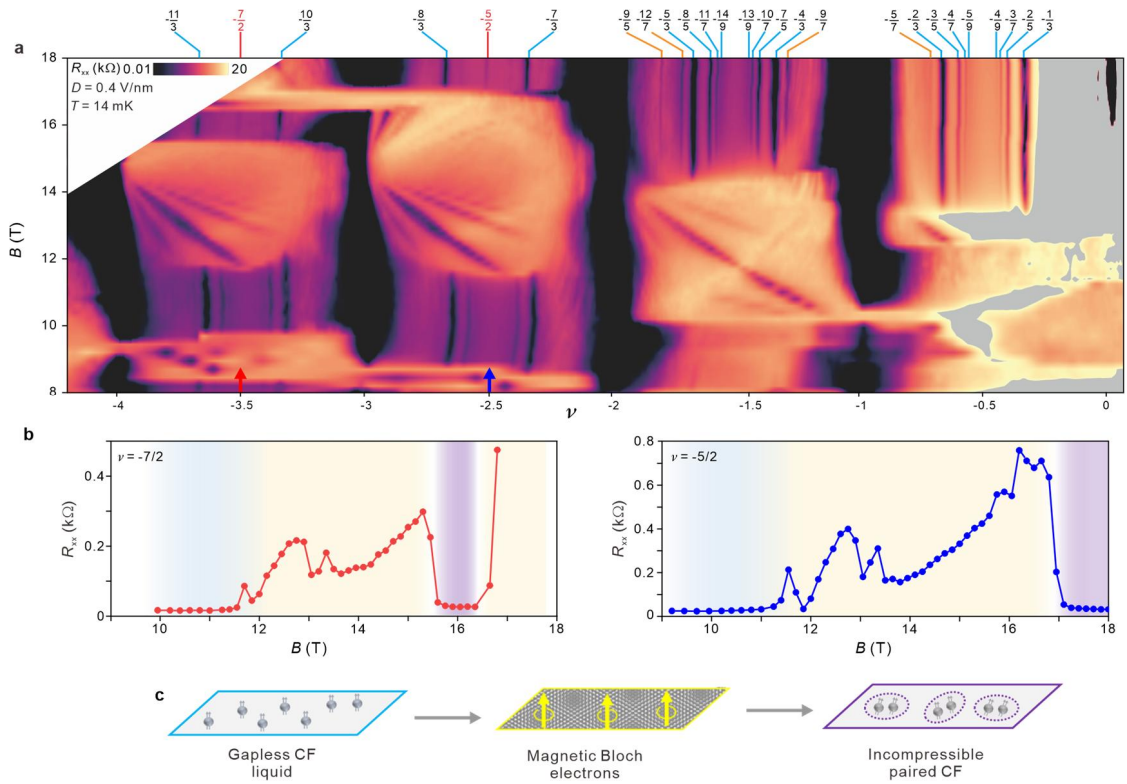

**Fig. 4 | Multiple phase transitions. a,** $R_{xx}$ as a function of filling $v$ and magnetic field $B$ at $T$ = 14 mK and $D$ = 0.4 V/nm. The Hofstadter butterfly pattern emerges within the region of Landau level crossings along tuning $B$-field. **b,** Linecuts of $R_{xx}$ along the red and blue arrows in **a**, revealing well-defined continuous phase transitions. The blue, yellow, and purple regions represent distinct phases, each associated with different quasiparticles, as depicted in **c**. **c,** Schematic illustrations of the distinct phases and their corresponding quasiparticles.

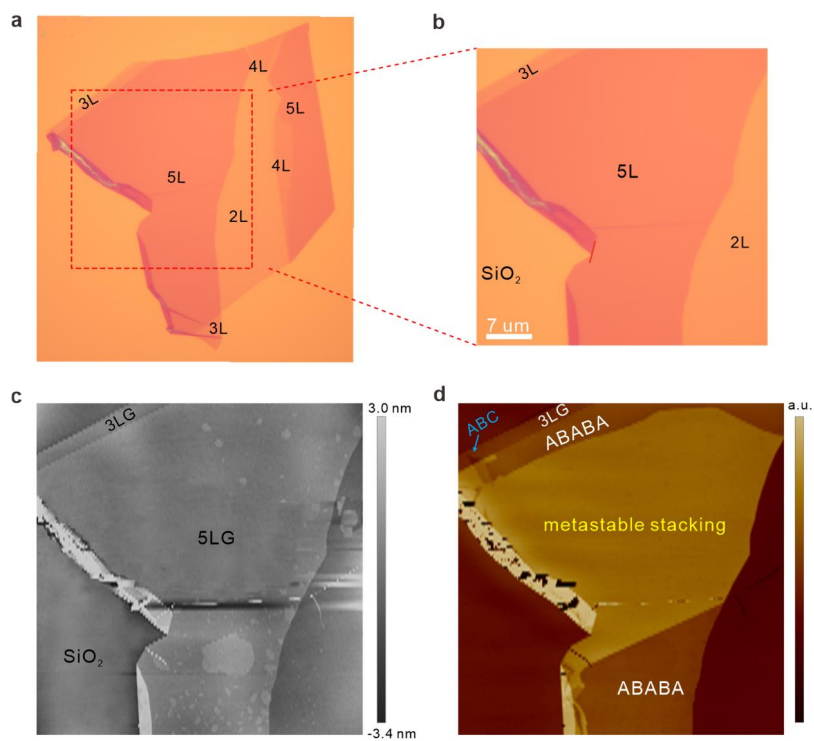

**Extended Data Fig. 1 | Imaging of ABCBC pentalayer graphene. a**, Optical image of the exfoliated graphene fleck on SiO$_2$/Si, where the layer number is identified by the reflection contrast between graphene and substrate using standard white-light microscopy (mainly the green channel). **b**. Zoom-in of the pentalayer (5L) region. **c**, Atomic force microscope (AFM) topography map of the same sample. **d**, Corresponding near-field infrared image. The brightest area is tentatively identified as a metastable stacking (e.g., ABCBC), though the exact structure remains uncertain from this characterization alone. The surrounding regions appear to be Bernal stacking (ABABA). Notably, a dark domain in the trilayer region is clearly visible and can be reliably assigned to ABC stacking based on prior experience. This domain extends into the pentalayer region, exhibiting a third distinct contrast that indicates the presence of an additional metastable stacking order.

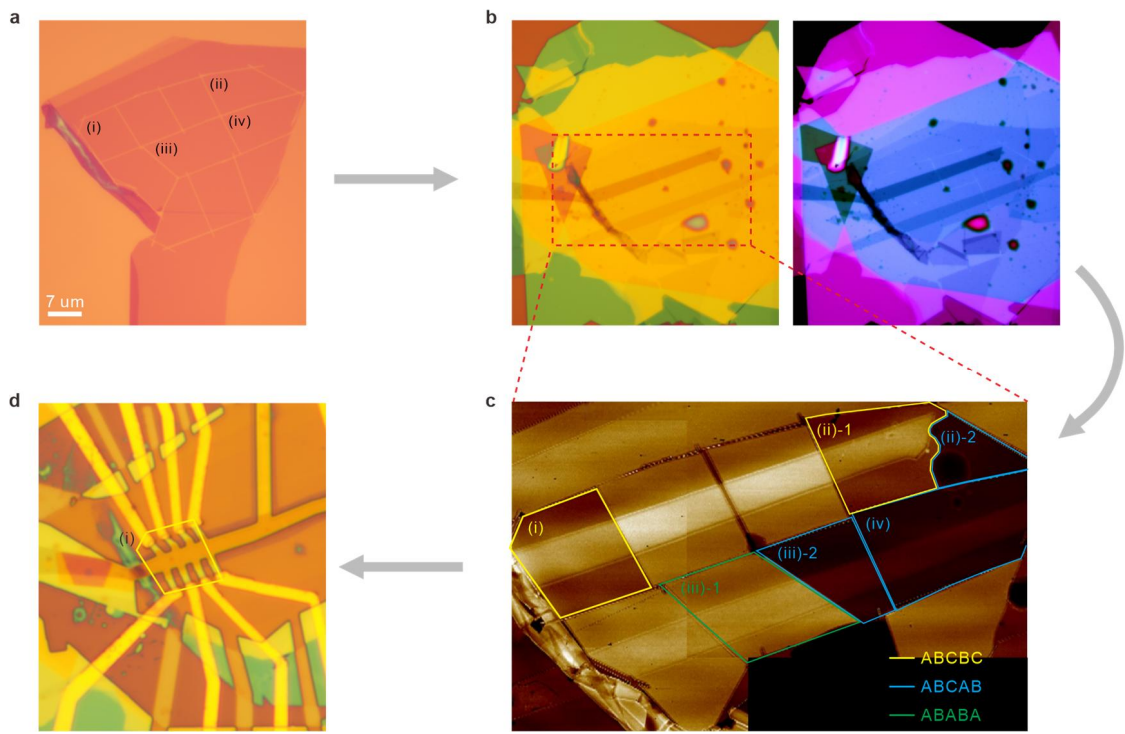

**Extended Data Fig. 2 | Fabrication of ABCBC pentalayer graphene device. a**, Optical image of pentalayer graphene after being cut into pieces using electrode-free anodic oxidation nanolithography. **b**, Corresponding optical images after ABCBC is encapsulated with hBN on both sides and bottom graphite gate. The right-side picture is a high contrast version, for a better visibility. **c**, Identification of ABCBC in the hBN encapsulated structure using the phonon-polariton assisted scanning near-field microscopy. As shown in the image, ABCAB shows the darkest contrast (outlined in blue), ABCBC is in the middle contrast (yellow) and ABABA (green) is the brightest. The regions labeled with (i), (ii), (iii), (iv) are the same as those in **a**. Notably, after transferring, parts of (ii) and (iii) changed to another stacking, denoted as (ii-1/2) and (iii-1/2) . Two bottom graphite gates also exhibit certain contrast inside the pentalayer graphene flakes. **d**, Optical image of the final device (region (ii)) after the nanofabrication process such as EBL, RIE and e-beam evaporation.

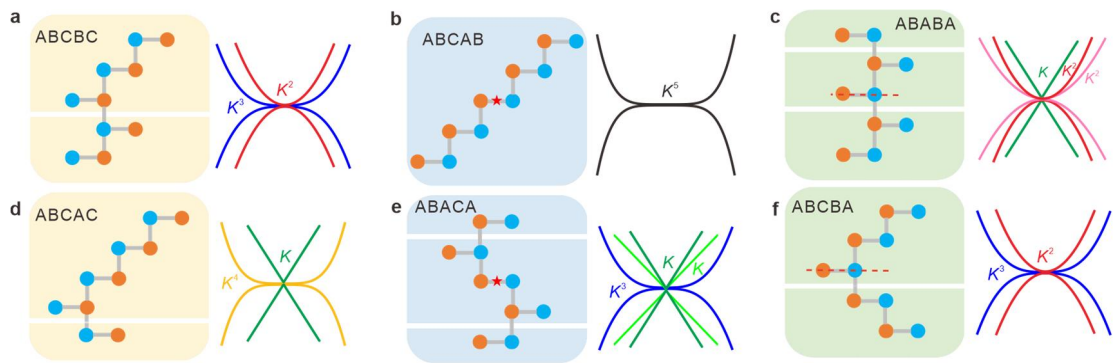

**Extended Data Fig. 3 | Possible stacking orders in pentalayer graphene. a-f**, Lattice and band structures of all six possible stacking orders of pentalayer graphene. Lattice structures with non-centrosymmetry, inversion symmetry and mirror symmetry are shaded in yellow, blue and green colors. Each lattice structure is decomposed into different blocks according to the chirality and the band structure can be simply obtained through the chiral decomposition.

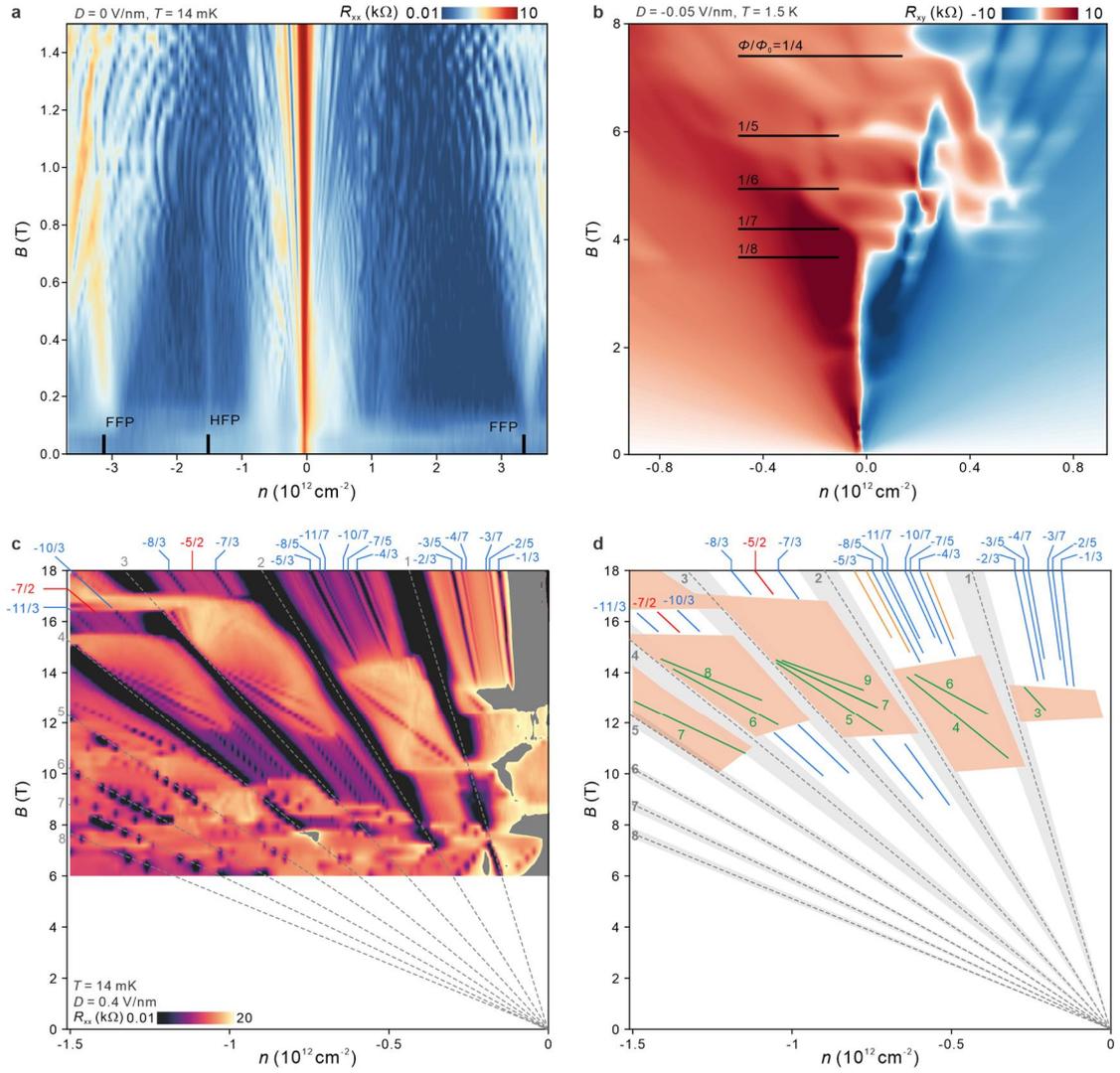

**Extended Data Fig. 4 | Landau fan diagram. a,** Low-field Landau fan diagram of $R_{xx}$ near $D = 0$. The full filling point (FFP) and half filling point (HFP) are marked, indicating the formation of a moiré superlattice between pentalayer graphene and hBN. **b,** $R_{xy}$ - $n$ - $B$ color plot when $D$ = -0.05 V/nm, $T$ = 1.5 K. Brown–Zak oscillations are visible and highlighted by the black solid lines. From the periodicity of these oscillations, a moiré wavelength of 12.8 nm is extracted. **c,** $R_{xx}$ - $n$ - $B$ color plot of ABCBC at $D$ = 0.4 V/nm when $T$ = 14 mK. **d,** Corresponding phase diagram concluded from **c**, the integer, even-denominator fractional and moiré-related quantum Hall states are indicated by grey, red and green lines. The odd-denominator fractional quantum Hall states with two and four flux quanta are indicated by blue and orange lines, respectively.

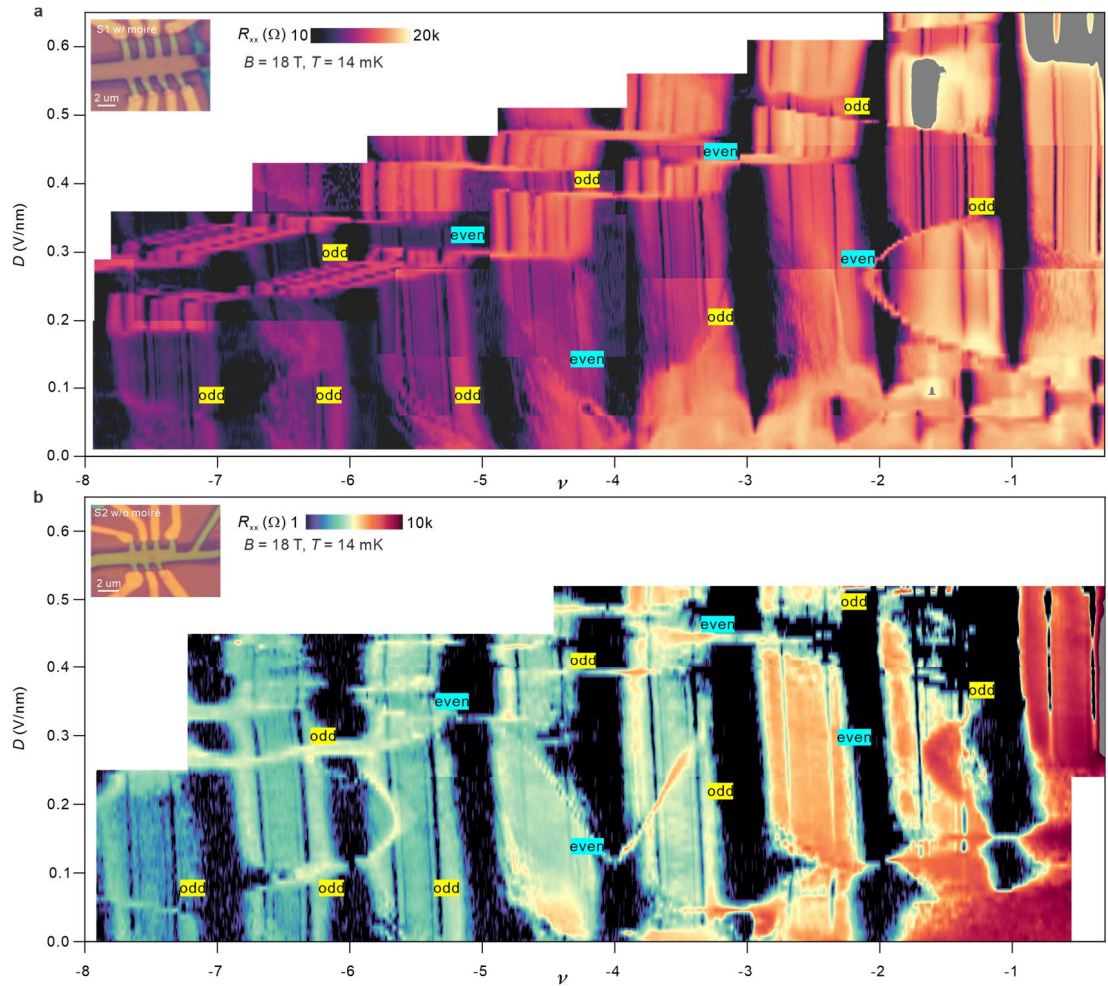

**Extended Data Fig. 5 | Comparison of phase diagram in ABCBC sample with (w/) and without (w/o) moiré. a,** Phase diagram of the sample S1 with moiré, reproduced from Fig. 4. **b,** Phase diagram of another ABCBC-5LG sample S2 without moiré, closely resembling that in **a**. Despite weaker $R_{xx}$ minima, a cascade of fractional quantum Hall features is still observed. Even and odd denominator FQH regions are marked in yellow and blue, respectively. Landau level crossings appear as increased resistance extending from the upper right to lower left, while no Bloch states are observed. We note that the suppression of adjacent Jain sequences confirms the emergence of half-filled states. These observations suggest that the effect of the moiré potential becomes prominent only when the Hofstadter gap is comparable to the Landau level gap near LL crossings.

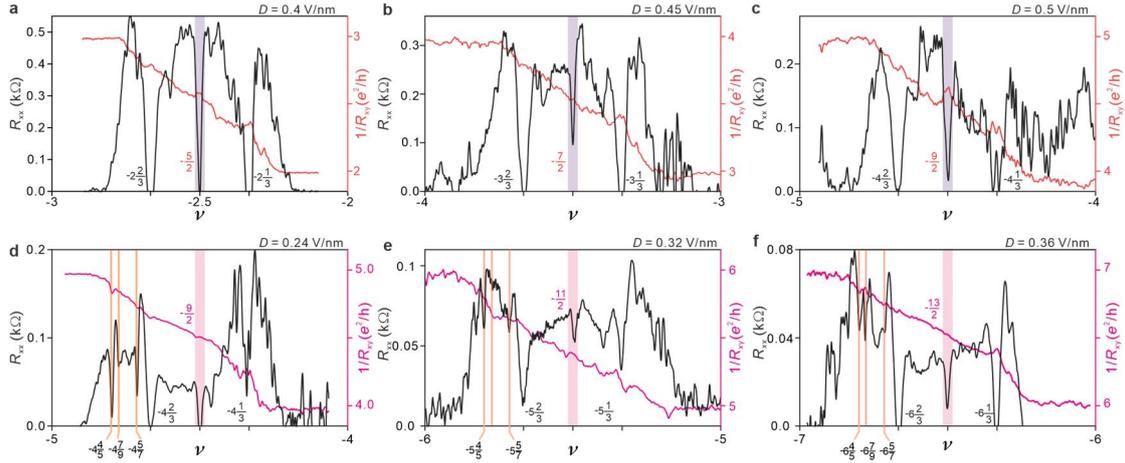

**Extended Data Fig. 6 | Even-denominator FQH states measured at $B$ = -18 T. a-f,** Linecuts of $R_{xx}$ and $1/R_{xy}$ within the filling ranges of $-3 \leqslant \nu \leqslant -2$ (**a**), $-4 \leqslant \nu \leqslant -3$ (**b**), $-5 \leqslant \nu \leqslant -4$ (**c** and **d**), $-6 \leqslant \nu \leqslant -5$ (**e**) and $-7 \leqslant \nu \leqslant -6$ (**f**), respectively. Compared to the data at +18 T (see Fig. 3), the signatures of the two sets of half-filled states remain robust and reproducible, including the accompanied $CF_4$ state in the second (pink shading) group. Remarkably, the $R_{xx}$ minima of the first state in each group, namely at -5/2 ($D$ = 0.4 V/nm) and -9/2 ($D$ = 0.24 V/nm), approach zero resistance in this case.

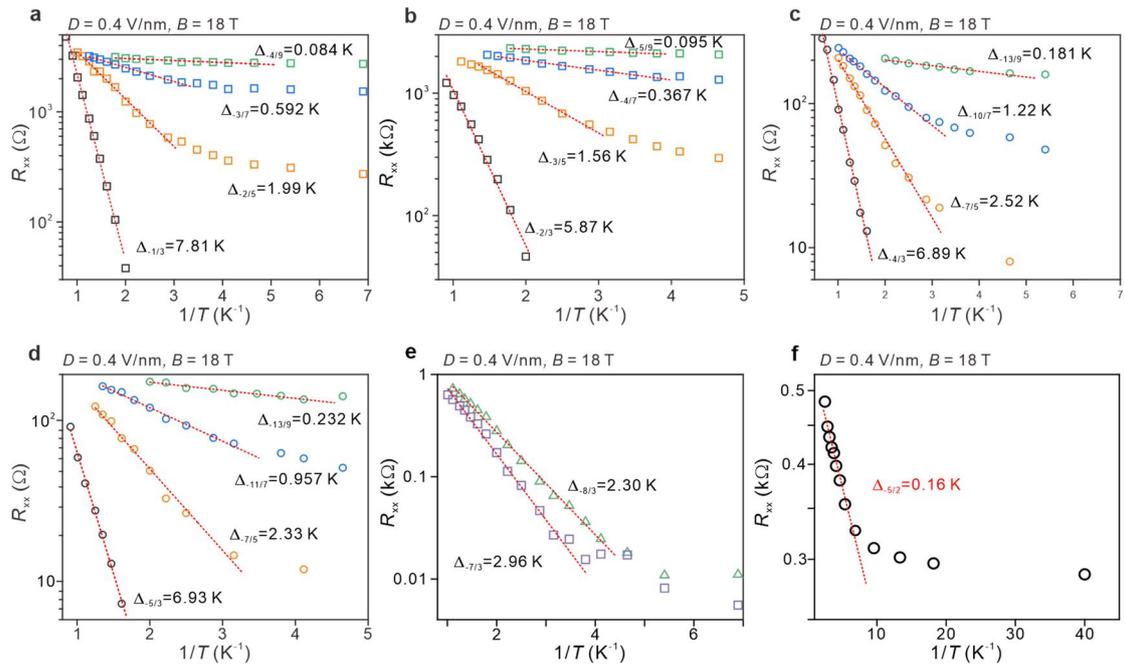

**Extended Data Fig. 7 | Thermal activation gaps in fractional quantum Hall states. a~f**, $R_{xx}$ versus $1/T$ plot of different fractional quantum Hall states when $D = 0.4$ V/nm and $B = 18$ T, the $R_{xx}$ axis is in the log scale, the labeled thermal activation gap is fitted using Arrhenius plot.

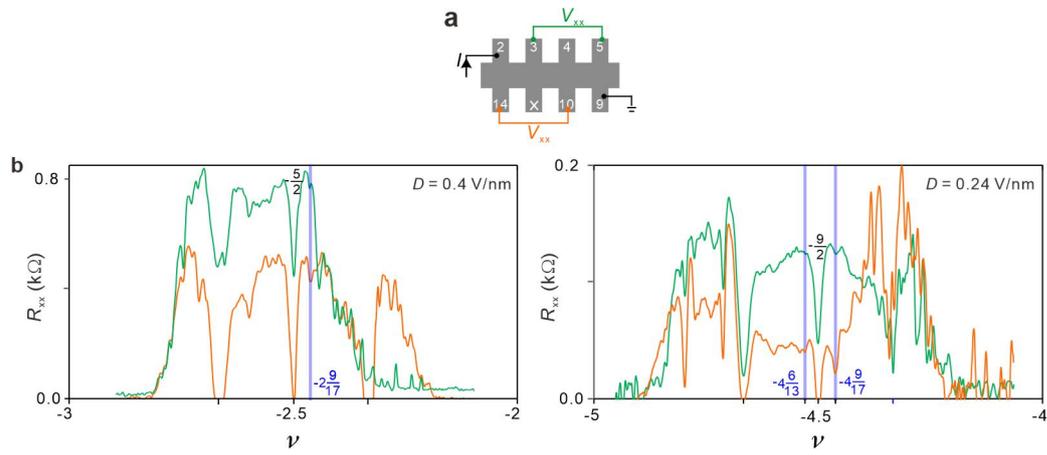

**Extended Data Fig. 8 | Particle-hole asymmetry at *v* = -5/2 and -9/2. a**, Different measurement configurations on two edges. **b.** $R_{xx}$ versus *v* plot in the vicinity of *v* = -5/2 and -9/2, respectively. Green and orange curves correspond to the two configurations in **a**. Weak but reproducible $R_{xx}$ minima are observed at $\tilde{v}$ = -6/13 and/or -9/17, marked by blue lines, which are the first daughter state of anti-Pfaffian state. Here, $\tilde{v}$ denotes *v* + 2 or *v* + 4.